\documentclass[12pt,reprint,aps,groupedaddress,showpacs,nofootinbib,prd]{revtex4-1}
\usepackage{amsmath,amssymb,mathrsfs,graphicx,slashed}
\usepackage{hyperref}
\usepackage{setspace}

\newcommand{\tGW}{t_{\text{GW}}}

\newcommand{\bv}{\boldsymbol{v}}

\newcommand{\PT}{\Phi^{T}}
\newcommand{\Mc}{M_{\text{comp}}}

\newcommand{\sm}{\sigma_{\text{max}}}
\newcommand{\Oo}{\Omega_{\text{orb}}}
\newcommand{\tC}{t_{\text{C}}}
\newcommand{\Bn}{B_{16}}
\newcommand{\nun}{\nu_{300}}

\newcommand{\Mn}{M_{1.4}}
\newcommand{\Rn}{R_{13}}
\newcommand{\an}{a_{7}}
\newcommand{\tB}{t_{B}}
\newcommand{\Tn}{T_{90}}
\newcommand{\bb}{\boldsymbol{B}}

\begin{document}

\title{Precursor flares of short gamma-ray bursts from crust yielding due to tidal resonances in coalescing binaries of rotating, magnetized neutron stars}

\author{Arthur G. Suvorov}
\email{arthur.suvorov@tat.uni-tuebingen.de}
\author{Kostas D. Kokkotas}
\affiliation{Theoretical Astrophysics, IAAT, University of T{\"u}bingen, Germany}


\date{\today}

\begin{abstract}
 
As evidenced by the coincident detections of GW170817 and GRB 170817A, short gamma-ray bursts are likely associated with neutron star-neutron star merger events. Although rare, some bursts display episodes of early emission, with precursor flares being observed up to $\sim 10$ seconds prior to the main burst. As the stars inspiral due to gravitational wave emission, the exertion of mutual tidal forces leads to the excitation of stellar oscillation modes, which may come into resonance with the orbital motion. Mode amplitudes increase substantially during a period of resonance as tidal energy is deposited into the star. The neutron star crust experiences shear stress due to the oscillations and, if the resonant amplitudes are large enough, may become over-strained. This over-straining can lead to fractures or quakes which release energy, thereby fueling precursor activity prior to the merger. Using some simple Maclaurin spheroid models, we investigate the influence of magnetic fields and rapid rotation on tidally-forced $f$- and $r$- modes, and connect the associated eigenfrequencies with the orbital frequencies corresponding to precursor events seen in, for example, GRB 090510.

\end{abstract}


\maketitle

\section{Introduction}

It has long been thought that short gamma-ray bursts (SGRBs) are associated with neutron star-neutron star (NSNS) merger events \cite{sgrb1,sgrb2,sgrb3}. This proposal is strongly supported by the combined detection of the gravitational wave (GW) event GW170817 from coalescing NSs and the subsequent SGRB that was observed by Fermi and INTEGRAL $\sim 1.7$ s after \cite{grbgw,grbgw2}. Although not observed for this particular event, some SGRBs are preceded by `precursor' flares: energetically weaker but spectrally similar flashes are seen several seconds prior to the main burst in some cases \cite{koshut95,troja10,min17,zhong19}. If the main episode arises within $\lesssim 2$ seconds after the merger, this would imply that some precursors occur before coalescence. Physical parameters inferred from precursor flare measurements, such as the strength of the mutual tidal strain on the progenitors at the orbital frequencies to which the flashes correspond, may therefore carry information about fundamental properties of the progenitor NSs, such as their equation of state (EOS) \cite{kokk95,tanj10,rezz14,gw17ns}.

In general, a perturbed NS oscillates as a superposition of modes, the amplitudes of which decay gradually due to the emission of GWs \cite{ipslin90,kokk99}. Pulsation modes are typically characterised according to the nature of their restoring force. For example, the fundamental $f$-modes are primarily restored by the hydrostatic pressure, while the dominant restoring force for the inertial $r$-modes is the Coriolis force \cite{unno79}. Fluid elements are displaced by these oscillations, thus resulting in shear stresses being applied to the NS crust. If the mode amplitudes are large enough to over-strain the crust to the point that it breaks in some sense \cite{chug10,chug18}, energy may be released from the star in the form of quakes or cracks \cite{pons11,lander15,suvkok19}.

A star, as part of a binary system, experiences an external tidal force with a driving frequency that is proportional to the orbital frequency (e.g. \cite{zahn77}). The tidal potential, in addition to adjusting the oscillation spectrum \cite{cowling41,denis72,gold1,gold2} induces a net quadrupole moment inside the star. This quadrupole also leads to shearing, though crustal failure due to tidal stresses alone likely only occurs within the final $\lesssim 10^{2} \text{ ms}$ of the inspiral \cite{penner11}. However, at certain orbital separations, a particular mode eigenfrequency may match the driving frequency, thus bringing the mode into resonance for a period of time, during which tidal energy is rapidly absorbed \cite{alex87,lai94,gold94,quat,lai97}. The absorbed energy over a resonance time-scale translates into a maximum mode amplitude, which, if greater than the critical number necessary to instigate crustal failure, leads to the consideration that crust yielding due to resonant mode excitations may be responsible for SGRB precursors \cite{tsang12,tsang13}. 

In this paper, we investigate tidally-forced oscillations and the induced crustal strains of rotating, magnetised NSs, to then compare quake energetics and resonance times with the luminosities and orbital frequencies observed for precursor flares from SGRBs \cite{koshut95,troja10,min17,zhong19}. Strong magnetic fields may be important in this scenario, as there is some evidence to suggest that those binaries which emit precursor flares contain magnetars. {For example, the majority of precursor flares exhibit a non-thermal spectrum \cite{zhong19}, which would be expected if Alfv{\'e}n waves propagating along open field lines are the primary means of the associated energy transport, though this requires a high surface field strength $B \gg 10^{13} \text{ G}$ \cite{thom95,tsang12}.} Recent estimates (between $\sim 0.4\%$ \cite{min17} and $\lesssim 2.7\%$ \cite{zhong19}) for the proportion of SGRBs that show precursor activity are also consistent with magnetar birth rates expected from population synthesis models \cite{gull15}. Oscillating NSs also emit gravitational radiation \cite{mtw}, so precursors attributable to large mode amplitudes should be accompanied by appreciable GW signals \cite{kokster98,parisi18,kru19}, especially if the stars have intrinsic quadrupole moments from magnetic deformations \cite{mlm13,mast15,mmra11}, which may be detectable with existing and upcoming GW observatories \cite{ligogrb}.


In any case, we adopt simple (equilibrium) models of constant density stars, and build on the Maclaurin spheroid solutions to present an analytic approach which can also account for rapid rotation \cite{chandra69,lind99,brav14}. Although clearly not realistic, the leading-order expressions for $f$- and (generalised) $r$-modes are quantitatively similar between the Maclaurin spheroids and stars with more realistic EOS \cite{andkok98,mlo99,lau10} (see Sec. III). Furthermore, it has been suggested that certain universality relationships between NS parameters observed in simulations for differing EOS (such as those relating the moment of inertia, Love numbers, and quadrupole moments defining the ``I-Love-Q'' relations \cite{ilq1,ilq2}) stem from the fact that the spectrum of an incompressible star reasonably approximates that of a star with a realistic nuclear EOS \cite{ilq3,ilq4,ilq5}. The simple models presented here are meant to serve as a proof of concept, to see whether strong magnetic fields or tidally-induced spectrum shifts can alter the viability of the $f$- or $r$-mode tidal resonance and subsequent quake scenario to explain SGRB precursors. 


This paper is organised as follows. In Section II we briefly discuss SGRBs, the properties of the precursor flares observed in some cases, and how tidal effects and strong magnetic fields may be relevant. Section III introduces the Maclaurin spheroid solutions, recaps the theory of their pulsations, and investigates the maximum mode amplitudes achievable during resonance. Estimates for the mode frequency shifts due to tidal and magnetic forces are given in Section IV. Section V then assesses the relationship between the resonant mode amplitudes and those necessary for crust yielding, to compare the energy available via quakes with the luminosities of the observed precursors. Some discussion is offered in Section VI.

We adopt the following notation for compactness throughout: $\Bn = B/(10^{16} \text{ G})$, $\nun = \nu / (300 \text{ Hz} )$, $\omega_{3} = \omega / (10^3 \text{ Hz})$ (similarly for frequencies in the inertial frame $\omega_{i,3}$ or for `unperturbed' values $\omega_{0,3}$), $\Mn = M / (1.4 M_{\odot})$, $\Rn = R / \left(13 \text{ km} \right)$, and $\an = a / \left(10^{7} \text{ cm} \right)$, where the symbols will be defined when introduced.

\section{Short gamma-ray bursts}

The SGRB event GRB 170817A was preceded $1.7$ s earlier by the GW signal GW170817 from a coalescing NS binary \cite{grbgw,grbgw2}. Given the $\sim 40$ Mpc distance of the source, this coincident detection provides strong evidence for the long-thought hypothesis that SGRBs are associated with compact merger events involving NSs \cite{sgrb1,sgrb2,sgrb3}. Although GW astronomy is still in its infancy, the first GRB was detected over 50 years ago \cite{firstgrb}, and many statistical analyses of the latter events have since been performed. GRBs are typically categorised according to their duration $T_{90}$ (i.e. the time interval in which 90$\%$ of the total photon count is detected within the prompt emission), with short bursts having $T_{90} \lesssim 2$ s and long ones having $T_{90} \gtrsim 2$ s \cite{kouv93}.

\subsection{Precursor flares}

As first reported by Koshut and collaborators \cite{koshut95}, some SGRBs\footnote{Long GRBs also show precursor flare activity (which is in fact more common), but the early emissions are much weaker energetically and have softer spectra \cite{lgrb1,lgrb2}. Moreover, since most long GRB are thought to be associated with core-collapse supernovae in low metallicity environments \cite{woosley93}, we do not consider them in this paper.} are preceded by precursor flares that are somewhat less intense but phenomenologically similar to the main episodes. The aforementioned authors estimated that $\sim$ 3$\%$ of bursts within the Burst and Transient Source Experiment (BATSE) showed precursor activity, with a $3 \sigma$ correlation between the respective $T_{90}$ durations of the pre- and main bursts. This estimate concerning the number of SGRBs hosting precursors varies substantially in the literature, the main reason being that the identification of a precursor is highly sensitive to the actual definition of what constitutes pre-emission; stipulating that the time interval between the precursor and the main GRB need not exceed the $T_{90}$ of the main burst, the optimistic estimate that $\lesssim 10\%$ of bursts may admit precursors was obtained in Ref. \cite{troja10}. A more recent analysis involving a study of $519$ SGRBs, which enforced the above condition, concluded that only $\lesssim 0.4\%$ of bursts show precursor activity \cite{min17}, while the analysis of Ref. \cite{zhong19}, which did not enforce this condition, found that $18$ of the $660$ bursts $(\lesssim 2.7\%)$ within their sample exhibited precursors, though these latter authors also included events of lower significance $(\sigma \lesssim 2)$.

In Table \ref{tab:sgrbdata} we present relevant properties for the most statistically significant SGRB precursor candidates discussed in the above references. Denoting the time of the pre-emission relative to the main burst by $t_{B} - t$, we see that GRBs 100717, 130310, and 071030 showed precursor activity a few $\sim$ seconds prior to the main burst ($t_{B} - t \lesssim 4.5$ s), while other events exhibited precursors much closer to the main emission, $t_{B} -t \lesssim 1$ s. The durations of these pre-emission episodes, in line with the initial findings of Ref. \cite{koshut95}, are correlated with the respective time lags. GRB 090510 is exceptional in the sense that two precursor events were identified \cite{troja10}, one occurring $\sim 13$ s prior, and a second occurring $\sim 0.5$ s prior (which lasted almost until the main burst). The final column of Tab. \ref{tab:sgrbdata} shows the inferred (Keplerian) orbital frequency $\Omega_{\text{orb}}$,
\begin{equation} \label{eq:kepfreq}
\Oo = \sqrt{ \frac {G \left( M + \Mc \right)} {a^3} },
\end{equation}
where $M$ and $\Mc$ are the masses of the binary stars with orbital separation\footnote{We ignore effects related to the eccentricity of the orbit; for the last stages of binary inspiral with $a \lesssim 10^{2} R$, angular momentum losses due to gravitational radiation tend to circularise the orbit \cite{peters,peters2}, so the orbital separation and semi-major axis are likely to roughly coincide. However, highly eccentric orbits may present more opportunities for resonances in general \cite{alex87,quat}.} $a$, to which the relative time lag corresponds in the following sense.

Due to the emission of gravitational radiation, $a$ decays with time. In particular, matching the GW luminosity with the rate of change of the orbital energy yields the well-known equation for $a(t)$ \cite{mtw},
\begin{equation}
\dot{a} = - \frac {64 G^3} {5 c^5} \frac {M^3 q \left( 1 + q \right)} {a^3},
\end{equation}
for mass-ratio $q = M_{\text{comp}} / M$, which has solution
\begin{equation} \label{eq:orbdecay}
a = \frac {\left[ 81 c^5 R^4 - \tfrac {256} {5} G^3 M^3 q \left( 1 + q \right) \left( t - \tC \right) \right]^{1/4} } {c^{5/4}},
\end{equation}
where $\tC$ is the coalescence time, occurring when $a \lesssim 3 R$ for (averaged) stellar radius $R$ \cite{ho99}. If we assume that $t_{B} \approx \tC$ (see below), expression \eqref{eq:orbdecay} allows us to estimate the orbital separations and frequencies at the times when the precursor events took place, which will be necessary to match with the mode frequencies in the resonance scenario (see. Sec. III. B). For GRB 090510, for example, expression \eqref{eq:orbdecay} gives us that $\Omega_{\text{orb}}\left(a\left(t-\tC=0.5 \right) \right) = 541.1$ Hz and $\Omega_{\text{orb}}\left(a\left(t-\tC=13 \right) \right)  = 160.5$ Hz.

\begin{figure}
\includegraphics[width=0.493\textwidth]{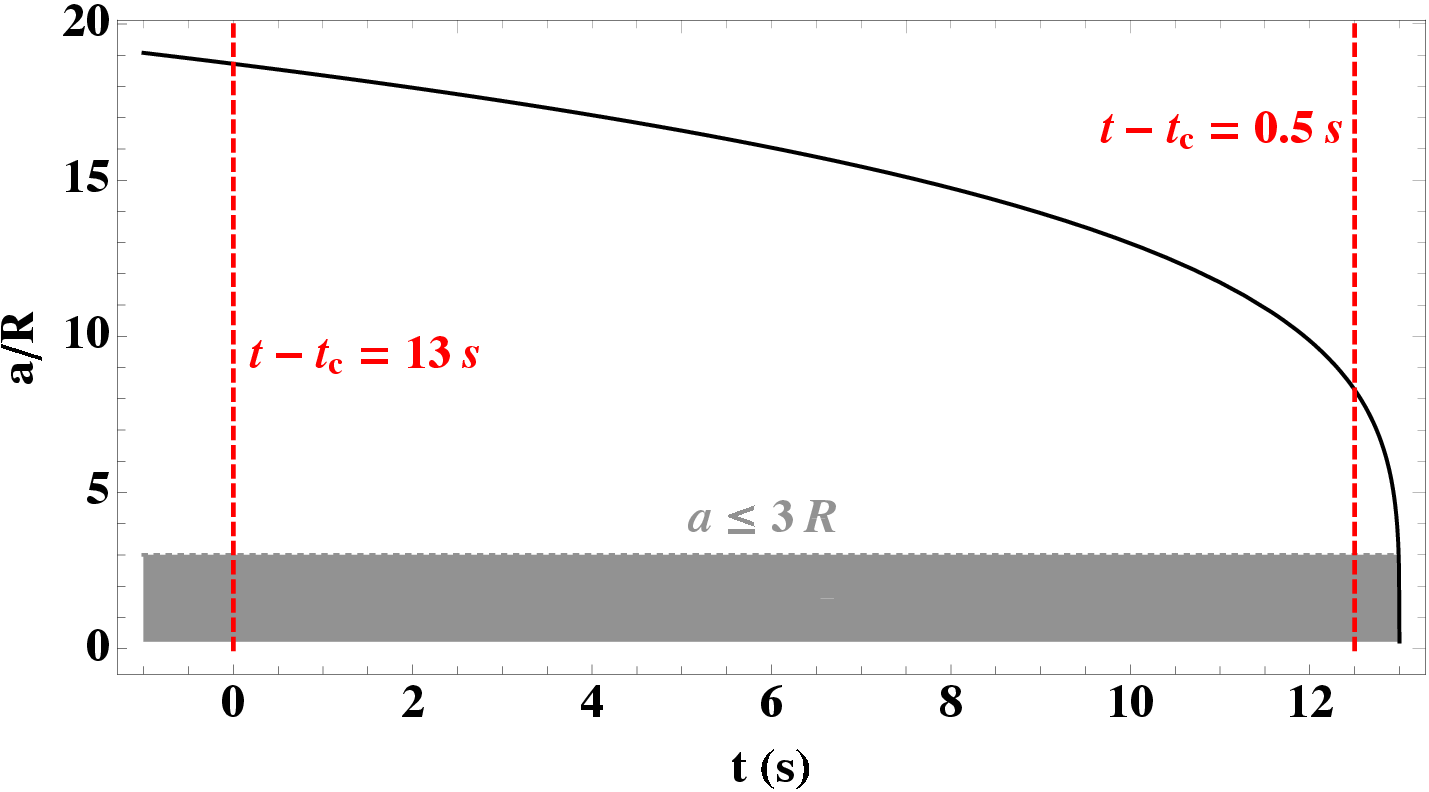}
\caption{Decay of the orbital separation $a$, normalised by stellar radius $R = 13 \text{ km}$, as a function of time for a binary with $M = \Mc = 1.4 M_{\odot}$. The vertical dashed lines mark $0.5$ s and $13$ s prior to coalescence (occurring when $a \lesssim 3 R$ \cite{ho99}), which, assuming that the main burst occurs at coalescence $t=\tC$, are the times at which precursor flares for GRB 090510 were identified \cite{troja10}. \label{orbsep}}
\end{figure}

Figure \ref{orbsep} shows the orbital separation \eqref{eq:orbdecay} as a function of time for an equal mass binary $M = \Mc = 1.4 M_{\odot}$, where coalescence occurs at $\tC = 13$ s. Precursor times for GRB 090510 are shown by the vertical dashed lines, measured as $13$ s and $0.5$ s prior to the main burst \cite{troja10}, corresponding to orbital separations $a \approx 19 R$ and $a \approx 8 R$, respectively. The grey shaded region represents the actual merger event, occurring at $a \lesssim 3 R$ \cite{ho99}.


In reality, the main burst will not occur at the instant of coalescence since there will be some non-zero time-scale associated with jet formation, which depends on the physics of the post-merger remnant. For instance, the mass of the remnant may exceed the Tolman-Oppenheimer-Volkoff limit but resist collapse due to strong differential rotation (the hypermassive NS scenario \cite{morrison04}). The SGRB \cite{sgrbh,hmnsitself} and X-ray afterglow \cite{hmnsitself2,plerion} may then be powered either by the NS itself or by a black hole formed through delayed collapse due to eventual angular momentum losses from GW emission (see Ref. \cite{grbreview} for a recent review). If the jet forms through highly-magnetised winds from the hypermassive NS, it is expected that $\tB-\tC \lesssim 100 \text{ ms}$ \cite{bern14} else neutrino emission from Urca cooling will choke the jet and produce a burst that lasts considerably longer than the $T_{90} \lesssim 2 \text{ s}$ defining SGRBs \cite{woos96,murg14}. If instead the hypermassive star collapses (or prompt black hole formation occurs), the magnetised accretion torus surrounding the freshly formed black hole may drive the jet via the Blandford-Znajek mechanism \cite{bz1,bz2}, and the jet may be launched within $\lesssim 30 \text{ ms}$ after collapse \cite{rezz11}.

In any case, treating the $1.7$ s delay time between GW170817 and GRB 170817A as canonical, (at least some of) the precursor flares described in Tab. \ref{tab:sgrbdata} were likely produced prior to the actual merger. In the next section, we explore the tidal resonance mechanism as a possible source of the precursors.

\begin{table*}
\caption{Properties of (the most significant) SGRB precursor candidates reported in Refs. \cite{troja10,min17,zhong19}. The associated orbital frequency \eqref{eq:kepfreq} is computed assuming an equal mass binary with $M = \Mc = 1.4 M_{\odot}$. }
  \begin{tabular}{c|c|c|c|c}
  \hline
  \hline
Precursor Event & Duration [$\Tn$ (s)] & Time relative to main burst [$\tB - t$ (s)] & Significance ($\sigma$) & Orbital frequency [$\Omega_{\text{orb}}$ (Hz)] \\
\hline
GRB 090510 & $0.05 \pm 0.02$ & $0.45 \pm 0.05$ & $\lesssim 4.6$ & $562.5^{+24.9}_{-21.4}$ \\
GRB 090510 & $\lesssim 0.4$ & $13$ & $5.2$ & $160.5$ \\
\hline
GRB 100717 & $0.3 \pm 0.05$ & $3.3$ & $12.8$ & $268.1$ \\
\hline
GRB 130310 & $0.9 \pm 0.32$ & $4.45 \pm 0.8$ & $10$ &  $239.7^{+18.5}_{-14.4}$ \\
\hline
GRB 071030 & $\lesssim 0.7$ & $2.5$ & $6.3$ & $297.4$ \\
\hline
GRB 060502B & $\sim 0.09$ & $0.32$ & $6.1$ & $637.5$ \\
\hline
GRB 100213A & $\sim 0.44$ & $0.68$ & $11.1$ & $483.0$ \\
\hline
GRB 140209A & $\sim 0.45$ & $1.06$ & $13.9$ & $409.6$ \\
\hline
GRB 160726A & $\sim 0.08$ & $0.39$ & $10.2$ & $592.9$ \\
\hline
\hline
\end{tabular}
\label{tab:sgrbdata}
\end{table*}

\subsection{Tidal resonances}


In the co-rotating frame of a NS spinning with angular velocity $\Omega$, suppose that a mode with azimuthal number $m$ has frequency $\omega$ (see Sec. III. A for details). In the inertial frame, the mode frequency reads
\begin{equation} \label{eq:inertialfreq}
\omega_{i} = \omega - m \Omega.
\end{equation}
For a general binary system, the tidal potential $\PT$ acts like an external driving force with (inertial frame) frequency $2 \Oo$ \cite{zahn77}. As such, if, at some point during the inspiral, we have that
\begin{equation} \label{eq:resonance}
|\omega_{i}| \approx 2 \Oo,
\end{equation}
then the mode comes into resonance with the orbital motion. At a given resonance frequency, the GW inspiral time $\tGW$ can be estimated from expressions \eqref{eq:kepfreq} and \eqref{eq:orbdecay} as
\begin{equation} \label{eq:tgwx}
\tGW \approx \frac {\Oo} {\dot{\Omega}_{\text{orb}}} = 0.63 \left( \frac {\mathcal{M}} {1.22 M_{\odot}} \right)^{-5/3} \omega_{i,3}^{-8/3} \text{ s},
\end{equation}
where $\mathcal{M} = M q^{3/5} \left(1 + q \right)^{-1/5}$ is the `chirp' mass, and we have used the resonance condition \eqref{eq:resonance} to rewrite $\Oo$ in terms of $\omega_{i}$. The duration of a particular resonance $t_{\text{res}}$ can be approximated by the time-scale during which the tidal driving is phase-coherent with the mode \cite{lai94,tsang12}, $t_{\text{res}} \sim \sqrt{ \tGW / \Oo}$, which, from \eqref{eq:tgwx}, reads
\begin{equation} \label{eq:restime}
t_{\text{res}} \sim 3.6 \times 10^{-2} \left( \frac {\mathcal{M}} {1.22 M_{\odot}} \right)^{-5/6} \omega_{i,3}^{-11/6} \text{ s}.
\end{equation}
It is interesting to note that, for the precursors described in Tab. \ref{tab:sgrbdata}, $t_{\text{res}}$ is within an order of magnitude of the $T_{90}$ reported for the flashes. 

In general, a given $n$-mode (e.g. $f$-mode) can be described by the associated Lagrangian eigenvector $\boldsymbol{\xi}_{n}$, which defines the extent to which fluid elements are displaced by the oscillations induced by that mode (see Sec. III. A). The external force generated by $\PT$ excites the oscillation modes, and results in an amount of energy being transferred to the star during the inspiral at a rate \cite{zahn77}
\begin{equation} \label{eq:energytransfer}
\dot{E}_{n}^{T} = \int d^{3} x \rho \frac {\partial \boldsymbol{\xi}^{\ast}_{n}} {\partial t} \cdot \nabla \PT,
\end{equation}
where the asterisk indicates complex conjugation. A resonant excitation [which fixes the orbital separation $a$ relative to $\omega_{i}$ through \eqref{eq:resonance}] of a particular mode then increases the respective amplitude $\alpha_{n}$ of that mode \cite{lai94,gold94}. Noting that, to leading order, $\PT \propto r^2 Y_{22}$ for spherical harmonic $Y_{22}$ (see Sec. IV), it is convenient to introduce the so-called overlap integrals\footnote{Note that these are defined with respect to a specific normalisation, namely that $\int dV \rho \boldsymbol{\xi}^{\ast} \cdot \boldsymbol{\xi} = M R^2$ (e.g. \cite{tsang12}).} $Q_{n}$, defined as \cite{alex87,kokk95}
\begin{equation} \label{eq:overlap}
Q_{n} = \frac {1} {M R^2} \int d^{3} x \rho \boldsymbol{\xi}_{n}^{\ast} \cdot \nabla \left( r^{2} Y_{22} \right).
\end{equation}
In terms of these integrals, the maximum mode amplitude $\alpha_{n,\text{max}}$, achieved during a period of resonance via tidal energy absorption \eqref{eq:energytransfer}, is given by (see Sec. 6 of \cite{lai94} for a detailed derivation)
\begin{equation} \label{eq:modeamp}
\begin{aligned}
\alpha_{n,\text{max}} &\approx \frac {\pi Q_{n}} {32} \left( \frac {\omega_{i}^2 R^3} { GM} \right)^{-5/12} \left( \frac {R c^2} {G M} \right)^{5/4} \left( \frac {2 q^{3/5}} {1+q} \right)^{5/6} \\
&= 6.2 \times 10^{-3} \omega_{i,3}^{-5/6} \Mn^{-5/6} \left( \frac {Q_{n}} {10^{-3}} \right) \left( \frac {2 q^{3/5}} {1+q} \right)^{5/6}. 
\end{aligned}
\end{equation}

In addition to the energy deposit \eqref{eq:energytransfer}, tidal interactions also induce a torque onto the NS, which results in angular momentum transfer; $\Delta J_{\text{res}} \approx  4M R^2 \omega_{i} \alpha^2_{n,\text{max}} $ during a period of resonance \cite{lai94,lai97,fuller11}. This angular momentum can naturally spin-up the star. An upper bound on the spin-up can be obtained by assuming that $\Delta J_{\text{res}}$ contributes only to the uniform rotation of the star, so that $\Delta \Omega \sim \Delta J_{\text{res}}/I$ for moment of inertia $I = 2 M R_{e}^2 /5$ \cite{chandra69}. Thus, during a period of resonance, the spin frequency of the star achieves a maximum increase due to the tidal torque by a factor
\begin{equation} \label{eq:spinup}
\frac {\Delta \Omega} {\Omega} \lesssim 2.1 \times 10^{-4} \nun^{-1} \omega_{i,3}^{-5/3} \Mn^{-5/3} \left(\frac {R} {R_{e}} \right)^{-2} \left( \frac {Q_{n}} {10^{-3}} \right)^2 ,
\end{equation}
for an equal mass binary, which may be significant if $Q_{n} \gg 10^{-3}$. 

In general, oscillation modes {which are not damped out at the crust-core interface may} shear the NS crust, in the sense that the physical stellar surface is `breathing' and fluid elements shift to a degree which depends on the amplitude $\alpha_{n}$ of the oscillation mode \cite{lee05,chir19}. During this time-scale $t_{\text{res}}$, if the mode amplitude \eqref{eq:modeamp} reaches a sufficiently large value, the crust may be strained to the point where it can no longer respond elastically and thus possibly crack (see Sec. V) \cite{chug10,chug18}. In addition to mode-induced stresses, if the magnetic field (and its excited oscillation) is also sufficiently strong, then the strain may exceed the critical threshold due to Maxwell stresses alone \cite{lander15,suvkok19}. We offer some motivation for considering strong magnetic fields in the next section.

\subsection{Hints for strongly magnetised progenitors}

Although far from conclusive, there is some evidence to suggest that at least one of the NSs involved in a merger event, which releases a precursor to the SGRB, is highly magnetised. We summarise this evidence as follows.

\begin{itemize}

\item{Most recent estimates indicate that precursors flares are emitted in between $\sim 0.4\%$ and $\lesssim 2.7\%$ of SGRBs \cite{min17,zhong19}. This means that the NSs involved are likely to be unusual in some sense, whether this mean rapid rotation, strong magnetic field, or otherwise. Population synthesis models suggest that $\lesssim 1\%$ of NSs are born with (surface) magnetic field strengths $B \gtrsim 10^{15} \text{ G}$ \cite{gull15}, so that $\lesssim 2\%$ of any given pair would contain at least one strong-field NS. While the internal field may be considerably stronger than the surface field \cite{thomp92,suvgep16}, the low proportion of precursors amongst SGRBs may indicate that a strong magnetic field is a precondition for early emission. {Moreover, precursors occurring at larger distances tend to be energetically weaker \cite{zhong19}.}}

\item{Many precursor flares display a predominantly non-thermal spectrum \cite{zhong19}. A natural explanation for this could be that mechanical energy stored within an over-sheared crust is relieved through Alfv{\'e}n waves, which transport energy to fuel precursor activity \cite{thom95}. This situation requires a strong field ($B \gg 10^{13} \text{ G}$ \cite{tsang12}), else the generation of pair-photon cascades from mode-induced backreactions into $B$ will tend to thermalise the spectrum. Furthermore, as noted in Ref. \cite{tsang13}, the surface magnetic field strength limits the extent to which energy can be extracted from the crust,}
\begin{equation} \label{eq:maxL}
L_{\text{max}} \sim 10^{50} \left( \frac {\omega_{3}  |\boldsymbol{\xi}| } { \Rn} \right) B_{\text{surf,15}}^2 \Rn^2 \text{ erg s}^{-1} ,
\end{equation}
which can nevertheless readily account for precursor energetics if $B$ is large enough.

\item{As discussed in Sec. II. A, a candidate theory for the launching of the SGRB jet itself (and for powering X-ray afterglows \cite{plerion}) is through the formation of highly-magnetised winds in a (possibly hypermassive) millisecond magnetar \cite{bern14,grbreview}. Although the $\alpha$-$\Omega$ dynamo or Kelvin-Helmholtz instability may explain the emergence of an ultra-strong field (saturating at $B \lesssim 10^{17} \text{ G}$) in the remnant \cite{thomp92}, a flux conservation argument suggests that the post-merger object is more likely to be highly magnetised if the progenitor stars are magnetars; see Ref. \cite{ciolfi100}. As such, an ultra-strong field for the remnant would be expected in this case, thus supporting the viability of this central engine, even if a dynamo does not operate.}

\end{itemize}

Again, we emphasise that the above points are certainly not conclusive, but do hint that magnetar-level field strengths for the progenitors may be tied to precursor activity. 


\section{Maclaurin spheroids}

Stars which are uniformly rotating and of constant density fall into the class of Maclaurin spheroids, the equilibrium properties of which have been studied in detail by Chandrasekhar \cite{chandra69}. 

Although often introduced using cylindrical coordinates, we stick with spherical coordinates $(r,\theta,\phi)$ throughout for ease of presentation, the origin of which $(r=0)$ is set as the center of the primary star. Equilibrium fluid profiles for uniformly rotating stars are given as solutions to the Euler equation (e.g. \cite{land87})
\begin{equation} \label{eq:euler}
0 = \nabla \left( \frac {p} {\rho} + \frac {r^2 \Omega^2} {2}  - \Phi \right),
\end{equation}
where $p$ is the stellar pressure, $\rho$ represents the density, $\Phi$ is the gravitational potential, and the velocity profile $\boldsymbol{v}$ has components $v_{r} = v_{\theta} = 0$, $v_{\phi} = \Omega r \sin \theta$. In particular, the Maclaurin spheroids have constant density,
\begin{equation}
\rho = \frac {3 M} {4 \pi R^3},
\end{equation}
where we note that the stellar volume $V = \tfrac {4 \pi} {3} R^3$. The pressure is given by
\begin{equation} \label{eq:maclaurinpres}
\begin{aligned}
p =& \pi G \rho^2 \left( \zeta \cot^{-1} \zeta - 1 \right) \\
&\times \left[ r^2 \left( 1 + 2 \zeta^2 + \cos 2 \theta \right) -2 R^2 \zeta \left( \zeta + \zeta^3 \right)^{1/3} \right],
\end{aligned}
\end{equation}
where $G$ is Newton's constant and $\zeta$ is a parameter related to the angular velocity $\Omega$ through the transcendental equation
\begin{equation} \label{eq:omegareln}
\Omega^2 = 2 \pi G \rho \zeta \left[ \left( 1 + 3 \zeta^2 \right) \cot^{-1} \zeta -3 \ \zeta \right],
\end{equation}
which must be solved numerically for $\zeta$ given some value of $\Omega^2 / \rho$. The spherical limit $\Omega \rightarrow 0$ corresponds to $\zeta \rightarrow \infty$.

The stellar surface $S$, defined by the vanishing of $p$ \eqref{eq:maclaurinpres}, is determined through the expression
\begin{equation} \label{eq:stellarsurface}
0 = r^2 \left( 1 + 2 \zeta^2 + \cos 2 \theta \right) - 2 R^2 \zeta \left( \zeta + \zeta^3 \right)^{1/3}.
\end{equation}
In particular, the surface of the spheroid forms an equipotential for the sum of the gravitational and centrifugal potentials within \eqref{eq:euler}, which is why the shape of the star is uniquely determined by the rotation parameter $\zeta$ in \eqref{eq:stellarsurface}. The star has equatorial and polar radii given by
\begin{equation} \label{eq:equatorialrad}
R_{e} = R \left( \frac {\zeta^2 + 1} {\zeta^2} \right)^{1/6},
\end{equation}
and
\begin{equation} \label{eq:polarrad}
R_{p} = R \left( \frac {\zeta^2} {\zeta^2 + 1} \right)^{1/3},
\end{equation}
respectively, where we note that $R_{e} \geq R_{p}$, indicating that the star is oblate.

In general, the star can support a maximum rotation rate $\Omega \approx 5.3 \Mn^{1/2} \Rn^{-3/2} \text{ kHz}$ (corresponding to $\zeta \approx 0.39$), which is $\lesssim 20\%$ lower than a realistic break-up limit \cite{cook94}. However, spheroids which have $\zeta \gtrsim 0.72$ are secularly unstable \cite{chandra69}, so we consider $\Omega \lesssim 4.9 \Mn^{1/2} \Rn^{-3/2} \text{ kHz}$, though this corresponds to a very rapidly rotating star with spin frequency $2\pi\Omega = \nu \lesssim 775 \text{ Hz}$.




\subsection{Free mode structure}

Pulsations of Maclaurin spheroids were initially studied by Bryan \cite{kelvin}, and have since been revisited in more completeness by Braviner and Ogilvie \cite{brav14}. Since a detailed analysis of the mode structure can be found in the aforementioned references, we will merely present the results which are most important for our purposes.

In general, linear oscillation modes arise when the background equilibrium undergoes a time-dependent perturbation, where each fluid variable $\chi$ (e.g. $\rho, p, \Phi, \boldsymbol{v}, \dots$) is perturbed according to $\chi \mapsto \chi_{\text{eq}} + \delta \chi(r,\theta) e^{i \left( \omega t + m \phi \right)}$, where $\chi_{\text{eq}}$ is the equilibrium profile. Since the background is an oblate spheroid, the perturbation variables can be decomposed as sums of spheroidal harmonics, which introduces an additional `quantum-number' $\ell$ into the scheme\footnote{Traditionally, $r$-modes are introduced through the use of magnetic-type vector spherical harmonics with $\delta \boldsymbol{v} \propto \boldsymbol{Y}^{B}_{\ell, \ell}$ (e.g. \cite{kokster98}), so that the `classical' $r$-modes are called $\ell = m$ modes. This is the notation adopted here. Note, however, that this velocity profile induces a density deformation of the form $\delta \rho \propto Y_{\ell+1, \ell}$ (e.g. \cite{lind99}), which is why these classical $r$-modes are sometimes called $\ell=m+1$ modes, as in Ref. \cite{brav14}. In this sense, for example, the $\ell = 4$, $m =2$ inertial modes of \cite{brav14} are what we call the $\ell = 3$, $m = 2$ $r$-modes.}. 

For a given $\ell$ and $m$, the permitted values of the mode frequency $\omega$ are determined by the imposition of boundary conditions on the stellar surface $S$, which forms a total pressure node, i.e. $p + \delta p = 0$ there. Physically speaking, the perturbed stellar surface should form through the advection of fluid elements on the background stellar surface; the thermodynamic enthalpy, which vanishes on the background surface \eqref{eq:stellarsurface}, should also vanish on the perturbed surface of the oscillating star \cite{fried78,lock99}. This condition can be enforced by demanding that the Lagrangian pressure perturbation $\Delta p$ vanishes on $S_{\text{eq}}$, viz.
\begin{equation} \label{eq:bdycdn}
\Delta p \big|_{p=0} = 0.
\end{equation}
By definition, the Lagrangian pressure perturbation reads \cite{mlo99}
\begin{equation} \label{eq:lagp}
\Delta p =\delta p + \boldsymbol{\xi} \cdot \nabla p,
\end{equation}
where we have introduced the displacement vector $\boldsymbol{\xi}$ related to the perturbed velocity through the simple relation
\begin{equation} \label{eq:lagpert}
\delta \boldsymbol{v} = \dot{\boldsymbol{\xi}} = i \omega \boldsymbol{\xi}.
\end{equation}

For $\ell = |m|$, solutions $\omega$ to \eqref{eq:bdycdn} which don't vanish in the limit $\Omega \rightarrow 0$ correspond to pressure-driven modes, i.e. the $f$-modes. For $m>0$, the $f$-mode frequencies $\omega_{f}$ are given as solutions to
\begin{equation} \label{eq:fmodefreq}
0 = \frac {\omega_{f}^2} {\Omega^2} - \frac {2 \omega_{f}} {\Omega} -  \frac {2 \ell} {B_{\ell}(\zeta)} \left[ \frac {1 + \zeta \left( 1 - \zeta \cot^{-1} \zeta \right) B_{\ell} (\zeta) } { \left( 1 + 3 \zeta^2 \right) \cot^{-1}\zeta - 3 \zeta } \right],
\end{equation}
where the function $B_{\ell}$ contains the associated Legendre polynomials $P^{\ell}_{\ell}$ and $Q^{\ell}_{\ell}$ through
\begin{equation} \label{eq:bfunction}
B_{\ell}(\zeta) = \left( 1 + \zeta^2 \right) \left[ \frac {1} {Q^{\ell}_{\ell}(i \zeta)} \frac {d Q^{\ell}_{\ell}(i \zeta)} {d \zeta} - \frac {1} {P^{\ell}_{\ell}(i \zeta)} \frac {d P^{\ell}_{\ell}(i \zeta)} {d \zeta} \right]. 
\end{equation}
In the formal limit $\Omega \rightarrow 0$, one can show that we recover the usual Kelvin mode \cite{kelvin} expressions from \eqref{eq:fmodefreq}, i.e.
\begin{equation} \label{eq:kelvin}
\frac {\omega_{f}^2} {\pi G \rho} = \frac {8 \ell \left( \ell - 1 \right)} {3 \left(2 \ell + 1 \right)} + \mathcal{O}(\Omega).
\end{equation}

For modes with $\ell \geq |m|$, the fluid also admits inertial modes ($r$-modes), in the sense that there are eigenvalues $\omega$ solving \eqref{eq:bdycdn} which have vanishing frequency in the non-rotating limit $\Omega \rightarrow 0$. In this paper, we are interested in those $f$- and $r$-modes with the strongest couplings to the tidal potential (see Sec. IV. A). Assuming that the orbital motion lies in the equatorial plane\footnote{This assumption has the implication that the classical $r$-modes with $\ell = m = 2$ cannot be excited by the leading-order, quadrupolar tidal potential (see Sec. IV. A), essentially because of orthogonality relations between the magnetic-type vector spherical harmonics and the fact that coefficients of cross-terms in $\PT$ vanish when the spin-orbit inclination angle is zero; see Ref. \cite{laiwu06} for a detailed discussion. Also, again because of orthogonality between different $m$ harmonics, $r$-modes with $m < 2$ cannot be excited by the leading-order tidal potential. Thus the leading-order $r$-modes for our case are the $\ell=3$ modes. For binaries with a significant inclination angle, the $\ell = m =2$ and $\ell = 2,3$, $m =1$ $r$-modes would also be relevant.}, this corresponds to those with $m=2$ and lowest $\ell$. For $\ell = 3, m =2$, the (positive $\omega$) inertial modes have frequency \cite{lind99}
\begin{equation} \label{eq:rmodefreq1}
\omega_{r} = 1.23 \Omega + \mathcal{O}(\Omega^2).
\end{equation}
Note that the eigenfrequencies \eqref{eq:rmodefreq1} are negative in the inertial frame \eqref{eq:inertialfreq}, thus indicating that the modes are subject to the gravitational radiation (CFS) instability \cite{fm78,and98,morsink98}. 

\begin{figure}
\includegraphics[width=0.493\textwidth]{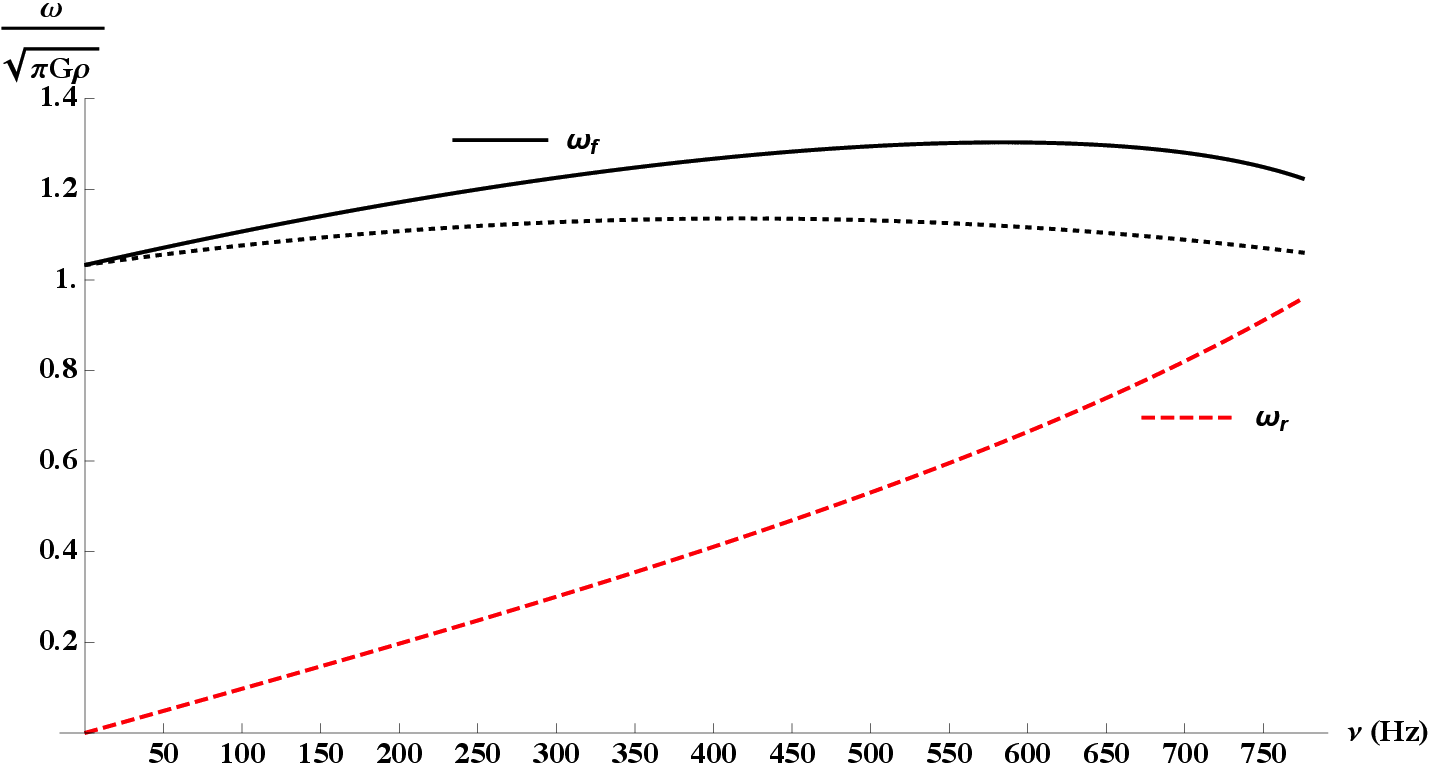}
\caption{$\ell = m = 2$ $f$- (black curves; the solid line shows the Maclaurin value while the dotted line shows the realistic EOS expression (21) of Ref. \cite{kokdon15}) and $\ell = 3, m =2$ $r$- (red, dashed curve) mode frequencies, normalised by the dynamical frequency $\sqrt{ \pi G \rho}$, as a function of spin frequency $\nu$. \label{fAndrFreq}}
\end{figure}

The (rotating frame) frequencies of the $f$- (black curves) and $r$- (red, dashed curves) modes, as functions of spin frequency $\nu$ up until the secular stability limit $\nu \lesssim 775 \text{ Hz}$, are shown in Figure \ref{fAndrFreq}. The solid $(f-)$ and dashed $(r-)$ curves indicate the Maclaurin spheroid eigenfrequencies. For contrast, the dotted curve shows the rotationally-corrected $f$-mode (see expression (21) of Ref. \cite{kokdon15}) eigenfrequencies determined in NSs with more realistic EOS. For the fundamental modes, we have that the rotational corrections to the frequencies disagree with the realistic values by at most $\approx 15\%$. Although calculations of the eigenfrequencies for $r$-modes with $\ell=3$ for realistic EOS are unavailable in the literature, Ref. \cite{mlo99} found similar deviations for the $\ell=2$ $r$-mode frequencies. We thus expect that the $\ell=3$ Maclaurin values reasonably approximate those in stars with more realistic EOS too (though general relativistic effects may be important \cite{kru19}). In general, with the exception of $f$-modes for very rapidly rotating stars with $\nu \gtrsim 600 \text{ Hz}$, the frequencies $\omega$ increase monotonically with $\nu$.


Finally, the perturbed versions of the Euler equations \eqref{eq:euler}, along with the continuity and Poisson equations, can be solved exactly to yield the perturbed fluid variables for the dominant  $f$- (i.e. $\ell = m =2$) and $r$- (i.e. $\ell = 3, m =2$) modes. 

The quantity of most importance to us is the Lagrangian displacement defined in \eqref{eq:lagpert}. For the $\ell = m = 2$ $f$-modes, one has \cite{brav14}
\begin{equation} \label{eq:fmodexi}
\boldsymbol{\xi}_{f} = 2 \alpha_{f} r \sin \theta e^{i \omega t} e^{2 i \phi} \{ \sin \theta, \cos \theta, i \},
\end{equation}
while for the $\ell = 3, m = 2$ $r$-modes, one finds the considerably more complicated expression
\begin{widetext}
\begin{equation} \label{eq:rmodexi}
\begin{aligned}
\boldsymbol{\xi}_{r} =& \frac {2 \alpha_{r} r \sin\theta} {R^2 \omega \Omega}  e^{i \omega t} e^{2 i \phi} \\
&\{2 \sin\theta \left( 42 r^2 \omega \left( \omega - \Omega \right) \cos^2\theta + \left( \omega + 2 \Omega \right) \left[ 3 b_{\zeta}^2 \left( \omega - 2 \Omega \right) + 7 i r^2 \left( \omega + \Omega \right) \sin^2\theta \right] \right) , \\
&2 \cos \theta \left( 3 b_{\zeta}^2 \left( \omega^2 - 4 \Omega^2 \right) + 7 r^2 \left( 3 \omega^2 \cos^2\theta + \left[ \left( i - 3 \right) \omega^2 + \left( 6 + 3 i \right) \omega \Omega + 2 i \Omega^2 \right] \sin^2\theta \right) \right), \\
& - i \left( -42 r^2 \omega^2 \cos^2\theta + \left( \omega + 2 \Omega \right) \left[ -6 b_{\zeta}^2 \left( \omega - 2 \Omega \right) + 7 r^2 \left( \omega + 4 \Omega \right) \sin^2\theta \right] \right) \},
\end{aligned}
\end{equation}
\end{widetext}
with 
\begin{equation}
b_{\zeta} = \frac{R} {\zeta^{1/3} \left( 1 + \zeta^2 \right)^{1/3}}  \sqrt{ \frac{ \omega^2 - 4 ( 1 + \zeta^2 ) \Omega^2 }  {\omega^2 - 4 \Omega^2} },
\end{equation}
mode amplitude(s) $\alpha$, and it is implied that only the real components are of interest. 

{It is important to note that in a realistic neutron star model with a solid crust, damping induced by viscous friction at the crust-core interface may prevent the modes from reaching the stellar surface \cite{bild99}. For a constant density model with $M = 1.4 M_{\odot}$ and $R = 10 \text{ km}$, spin frequencies $\nu \gtrsim \nu_{\text{crit}} \approx 39 \left( \mu / 10^{30} \text{ dyn cm}^{-2} \right)^{1/2}$ Hz for shear modulus $\mu$ are sufficient to ensure that $r$-modes strongly penetrate the crust \cite{lev01}.}


\subsection{Resonant amplitudes}

From the above expressions for $\boldsymbol{\xi}_{n}$, we can evaluate the overlap integrals \eqref{eq:overlap} to estimate the maximum mode amplitudes \eqref{eq:modeamp} achievable during resonance. For the $f$-modes, we find
\begin{equation} \label{eq:foverlap}
Q_{f} = 0.69 \frac {R_{e}} {R},
\end{equation}
in agreement with Ref. \cite{ho99}. For the $r$-modes, we have 
\begin{equation}
Q_{r} = 3.53 \times 10^{-3} \nun^2 \Rn^{-3} \Mn^{-1} + \mathcal{O}(\nu^4),
\end{equation}
as found in Ref. \cite{laiwu06}, where we note that higher order corrections in $\Omega$ are negligible except for very rapidly rotating models with $\nu \gtrsim 800 \text{ Hz}$. For an equal mass binary $(q=1)$, we therefore find that the resonant mode amplitudes \eqref{eq:modeamp} are
\begin{equation} \label{eq:famp}
\alpha_{f,\text{max}} \approx \frac {0.74 R_{e}} {R} \Mn^{-5/6} \left( \Mn^{1/2} \Rn^{-3/2} - 0.28 \nun \right)^{-5/6},
\end{equation}
and
\begin{equation} \label{eq:ramp}
\alpha_{r,\text{max}} \approx 1.68 \times 10^{-2} \nun^{7/6} \Mn^{-11/6} \Rn^{3}.
\end{equation}


Having introduced the oscillation modes of Maclaurin spheroids, we now turn to an investigation of how tidal (IV. A) and magnetic (IV. B) forces can modulate the mode frequencies shown in Fig. \ref{fAndrFreq}.

\section{Frequency modulations}

The introduction of a perturbing force $\delta \boldsymbol{F}$ into the Euler equations \eqref{eq:euler} leads to a modulation $\delta \omega$ in the mode frequencies, essentially because a change to the perturbed pressure profile leads to a shift in the eigenvalue solutions of equation \eqref{eq:bdycdn}. In general, these shifts are given by the exact expression \cite{unno79,bi13}
\begin{equation} \label{eq:freqpert}
\frac {\delta \omega} {\omega_{0}} = \frac {1} {2 \omega_{0}^2} \frac {\int dV \boldsymbol{\xi}^{\ast} \cdot \delta \boldsymbol{F}} {\int dV \rho |\boldsymbol{\xi}|^2},
\end{equation} 
where we denote the `unperturbed' mode frequencies found in the previous section as $\omega_{0}$. Equation \eqref{eq:freqpert} can be evaluated for some particular choices of $\delta \boldsymbol{F}$. 

\subsection{Tidal potential}
In addition to potentially exciting modes due to resonance \eqref{eq:resonance}, tidal forces also necessarily shift the mode frequencies \cite{cowling41,denis72}.  Treating the companion star as a point source, the tidal potential $\PT$ admits a multipole expansion of the form \cite{alex87,lai94,kokk95}
\begin{equation} \label{eq:tidalpot}
\PT = - \frac {G \Mc} {a} \left[ 1 + \sum_{k = 2} \left( \frac {r} {a} \right)^{k} P^{k}_{0}  \left(\cos \tilde{\phi} \sin \theta \right) \right],
\end{equation}
where $\tilde{\phi} = \phi - \phi_{c}$, with $\phi_{c}$ representing the angular position of the secondary star as measured from the perihelion of the orbit,
\begin{equation} \label{eq:orbfreq}
\phi_{c} = \Oo t,
\end{equation} 
with $\Oo$ given by expression \eqref{eq:kepfreq}. The leading-order $(k=2)$ term of $\PT$, most relevant for tidally-forced oscillations \cite{zahn77}, reads
\begin{equation} \label{eq:tidallead}
\PT = - \frac {1} {8} \frac {G \Mc} {r} \left( \frac {r} {a} \right)^{3} P^{2}_{2} (\cos \theta) e^{2 i \phi}  e^{i \lambda t},
\end{equation}
with $\lambda = 2 \Oo$ being the forcing frequency.  If the primary star is rotating with angular velocity $\Omega$, the forcing frequency in the co-rotating frame is obtained through $\Oo \mapsto \Oo - \Omega$.

The perturbing tidal force is thus given by
\begin{equation}
\delta \boldsymbol{F}^{T} = \rho \nabla \PT.
\end{equation}
We can now evaluate the frequency shift \eqref{eq:freqpert}. Since the Lagrangian displacements $\boldsymbol{\xi}$ and eigenfrequencies $\omega$ depend on the rotation rate in a complicated way (especially for the $r$-modes), we present, for convenience, best (least-squares) fits to the frequencies as functions of the stellar parameters.

For leading-order $f$-modes, we have
\begin{equation} \label{eq:tidalf}
\begin{aligned}
\frac{ \delta \omega^{T}_{f}} {42.3 \text{ Hz}} =& - q \left( \frac {\alpha_{f}} {0.1} \right)^{-1} \Mn^{3/2} \Rn^{3/2} \an^{-3} \\
&\times \left( 1 - 0.21 \nun + 0.055 \nun^2 \right),
\end{aligned}
\end{equation}
which, importantly, is negative for positive mode amplitude $\alpha_{f}$, and thus makes resonance `easier' to achieve in principle.  For $r$-modes, we have
\begin{equation} \label{eq:tidalr}
\begin{aligned}
\frac{ \delta \omega^{T}_{r}} {0.67 \text{ Hz}} =& q \left( \frac {\alpha_{r}} {0.01} \right)^{-1} \Mn^{2} \an^{-3} \\
&\times \left( 1 + 21.8 \nun^{-1} - 1.82 \nun \right) .
\end{aligned}
\end{equation}

Note that the shifts \eqref{eq:tidalf} and \eqref{eq:tidalr} are only, strictly speaking, valid near resonance \eqref{eq:resonance}. 

It is worth noting that, recently, a variation of the radiation-reaction secular instability (similar to the CFS instability \cite{fm78,and98,morsink98}) in tidally-forced NSs was shown to operate when $\Omega > \Oo$ \cite{pnig19}. The growth-time of this instability can be faster than the GW inspiral time \eqref{eq:tgwx} in the final $\lesssim 10$ s of inspiral, as relevant for precursors, and may thus further shift the mode eigenfrequencies. This will be investigated in future work.

\subsection{Magnetic field}

We begin by constructing an axisymmetric, dipolar\footnote{For simplicity, we consider only dipolar magnetic fields here, though it should be noted that multipolar components can introduce non-negligible deformations into the {hydrostatic} pressure profile for magnetar-like field strengths \cite{mlm13,mast15}.} mixed poloidal-toroidal magnetic field in the manner outlined in Ref. \cite{mmra11}, so that the Lorentz force can be introduced to evaluate \eqref{eq:freqpert}.

An axisymmetric magnetic field admits a Chandrasekhar decomposition \cite{c56} into poloidal and toroidal components, viz.
\begin{equation} \label{eq:bfield}
\bb = B_{0} \left[ \nabla \psi \times \nabla \phi + \left(\frac {E^{p}} {E^{t}} \frac {1 - \Lambda} {\Lambda} \right)^{1/2} \beta(\psi) \nabla \phi \right],
\end{equation}
where $B_{0}$ sets the characteristic field strength, $\psi = f(r) \sin^2\theta$ is a scalar streamfunction, and the toroidal component $\beta$ is a function of $\psi$ only. In expression \eqref{eq:bfield}, we have introduced the quantities
\begin{equation} \label{eq:epol}
E^{p} = \frac {1} {8 \pi} \int_{V} dV \left[ \left( \frac {1} {r^2 \sin\theta} \frac {\partial \psi} {\partial \theta} \right)^2 + \left( \frac {1} {r \sin\theta} \frac {\partial \psi} {\partial r} \right)^2 \right],
\end{equation}
and
\begin{equation} \label{eq:etor}
E^{t} = \frac {1} {8 \pi} \int_{V} dV  \frac {\beta(\psi)^2} {r^2 \sin^2\theta},
\end{equation}
which represent the poloidal and toroidal energies stored within the \emph{internal} magnetic field, respectively. In \eqref{eq:bfield}, $0 < \Lambda  \leq 1$ parameterises the relative strengths of the poloidal and toroidal components, e.g. $\Lambda = 0.5$ defines a field with an equal poloidal-to-toroidal energy ratio: $E^{p} = E^{t}$.

In general, many options are available for the function $f$ appearing within the streamfunction $\psi$, though we make the same choice as in Ref. \cite{mmra11},
\begin{equation} \label{eq:alphal}
f(r) = \frac {r^2} {8} \left[ 35 - 42 \left( \frac {r} {R}\right)^2 + 15 \left( \frac {r} {R}\right)^4 \right],
\end{equation}
which ensures that the magnetic field $\bb$ is finite everywhere and (approximately) continuous with respect to a current-free external field. We pick the function $\beta$ as
\begin{equation} \label{eq:betafn}
 \beta(\psi) =
\begin{cases}
(\psi - R^2)^2 / R^3&\textrm{for }\psi \geq R^2,\\
0&\textrm{for }\psi < R^2,
\end{cases}
\end{equation}
which ensures that the toroidal field is confined within an equatorial torus, as generally observed in time-dependent simulations (e.g. \cite{braith}). Magnetic field oscillations $\delta \bb$ are then determined as solutions to the Faraday equation,
\begin{equation} \label{eq:faraday}
0 = i \omega \delta \bb - \nabla \times \left( \delta \bv \times \bb + \bv \times \delta \bb \right),
\end{equation}
taking care to preserve the divergence-free condition $\nabla \cdot \delta \bb = 0$, where $\delta \bv$ is determined from \eqref{eq:lagpert} using the displacements $\boldsymbol{\xi}$ computed in the previous section.

\begin{figure}
\includegraphics[width=0.493\textwidth]{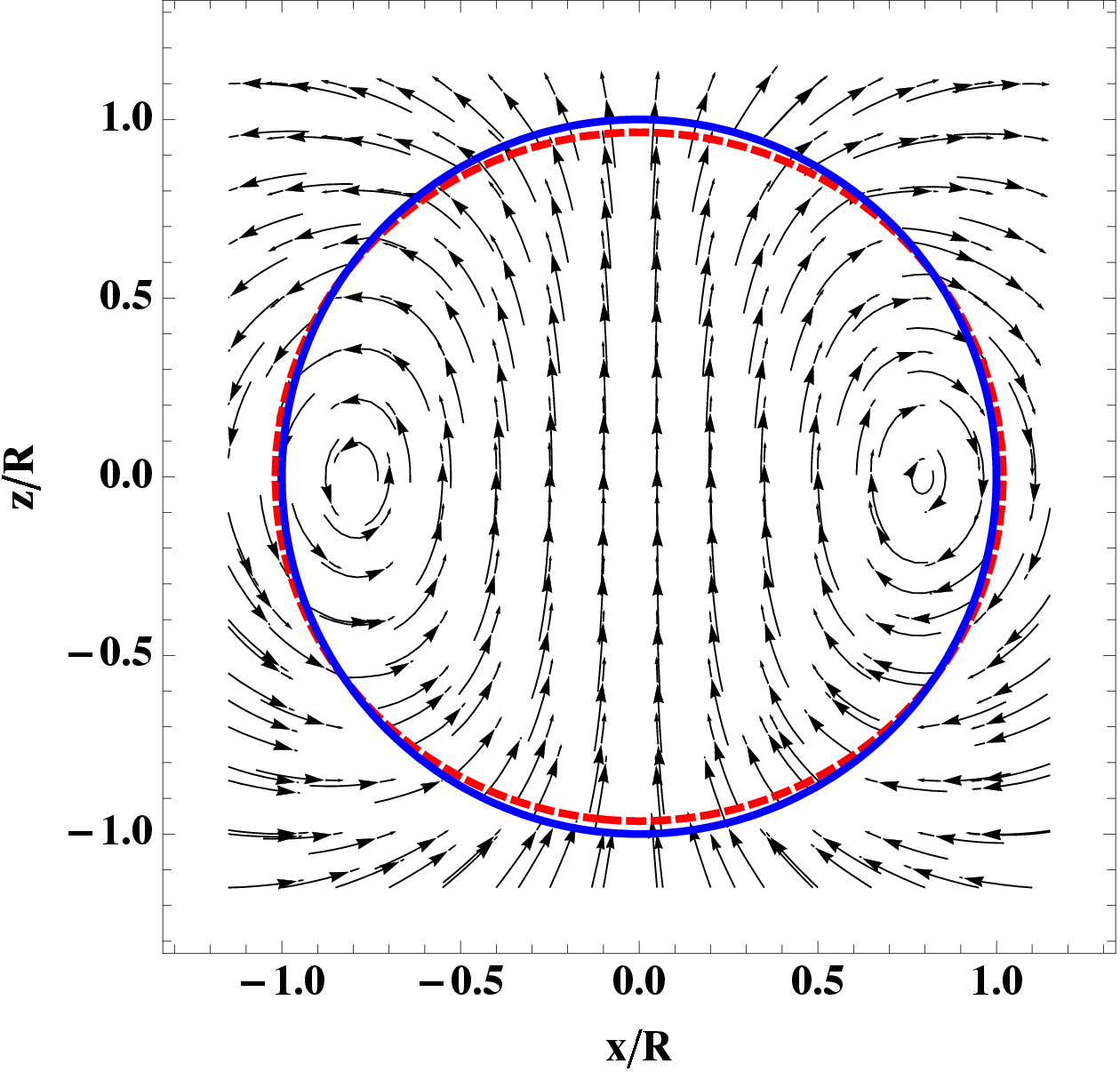}
\caption{Field lines for the background magnetic field $\boldsymbol{B}$ for a rotating configuration with $\nu = 300 \text{ Hz}$. Even at this spin frequency, the stellar surface is roughly spherical; the red, dashed curve shows the (true) stellar surface \eqref{eq:stellarsurface} while the blue curve shows the spherical surface $x^2 + z^2 = R^2$. \label{bgmagfields}}
\end{figure}

\begin{figure*}
\includegraphics[width=\textwidth]{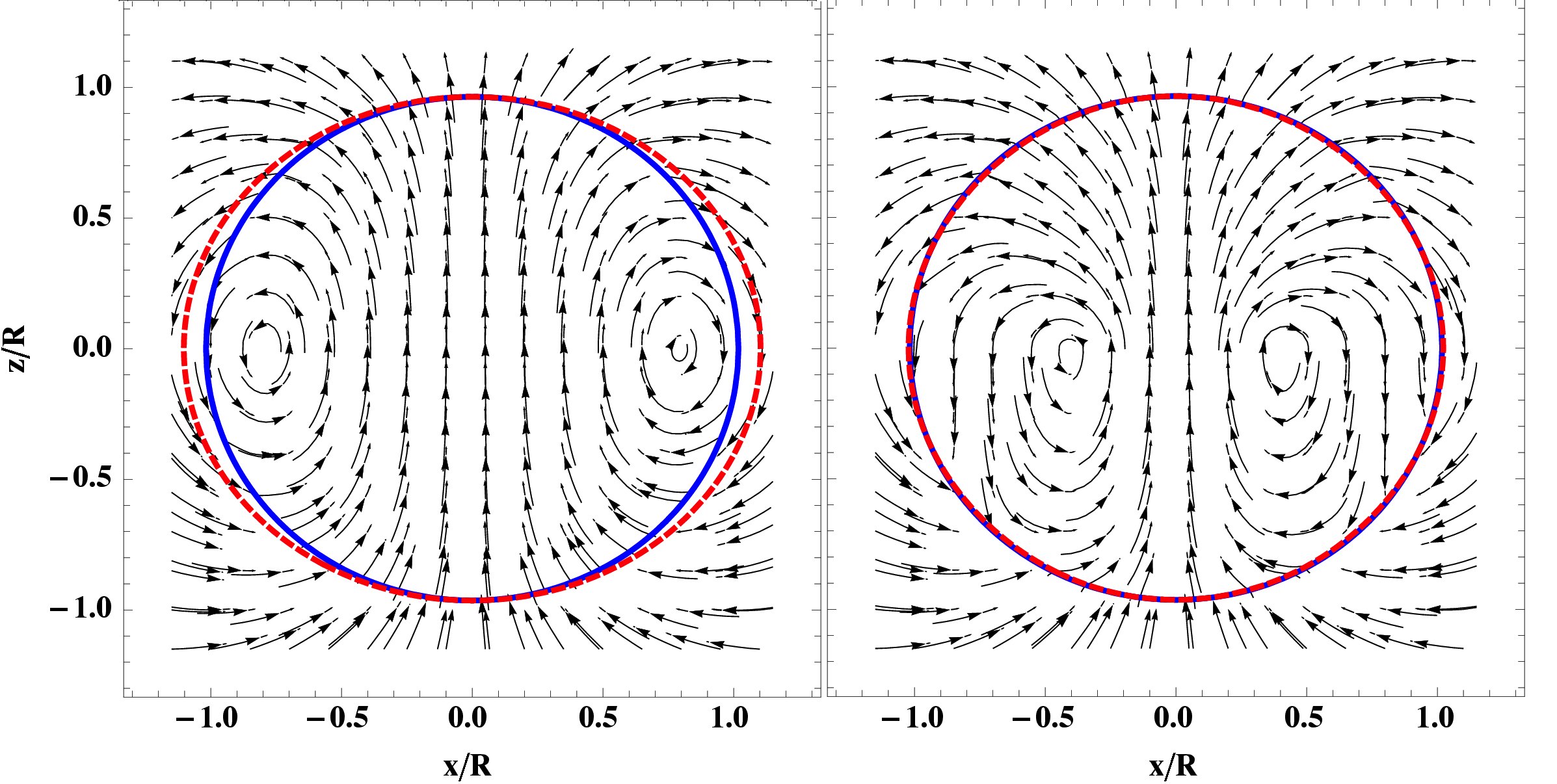}
\caption{Field lines for the total magnetic field $\boldsymbol{B} + \delta \boldsymbol{B}$ where $\delta \boldsymbol{B}$ is induced by $f$- (left panel; $\alpha_{f} = 0.04$) and $r$- (right panel; $\alpha_{r} = 0.02$) modes at $t=0$ for a rotating configuration with $\nu = 300 \text{ Hz}$. The field lines are laterally shifted (for $f$-) and twisted (for $r$-) relative to the background dipole configuration due to the mode perturbations. The red, dashed curve shows the \emph{perturbed} stellar surface, which is different for different mode perturbations, while the blue curve shows the background Maclaurin surface. \label{pertmagfields}}
\end{figure*}

Figures \ref{bgmagfields} and \ref{pertmagfields} show the magnetic field lines for the background $\boldsymbol{B}$ and total $ \bb + \delta \boldsymbol{B}$ configurations, respectively, for spin frequency $\nu = 300 \text{ Hz}$. In Fig. \ref{bgmagfields}, the blue curve shows the spherical surface $x^2 + y^2 = R^2$, while the red, dashed curve shows the (true) oblate surface defined by \eqref{eq:stellarsurface}, which is elliptical rather than spherical due to the centrifugal force in the Euler equation \eqref{eq:euler}. In Fig. \ref{pertmagfields}, however, the background Maclaurin surface is shown in blue, while the red, dashed curves show the (true) perturbed stellar surface at $t=0$, defined by the vanishing of the Lagrangian pressure perturbation $\Delta p$, which depends on the character of the perturbation through $\boldsymbol{\xi}$. In the latter Figure, we have that $\delta \bb$ is determined through \eqref{eq:faraday} for $f$- (left panel) and $r$- (right panel) mode perturbations. We see that the $f$-mode pulsations push the field lines laterally, thus shifting the toroidal geometry as the field lines are stretched due to the equatorial `breathing' typical for pressure-driven modes, which (quasi-)periodically make the star more oblate. On the other hand, the axial $r$-modes induce an approximately horizontal flow, which in turn winds the field lines and drags them along arcs directed towards the centre of the star. The stellar surface in this case is negligibly disturbed, as expected, since $r$-modes induce weak changes to the pressure profile. 

Having found $\delta \bb$ from \eqref{eq:faraday}, the perturbing Lorentz force,
\begin{equation} \label{eq:lorentz}
\delta \boldsymbol{F}^{B} = - \left(4 \pi \right)^{-1} \left[ \left(\nabla \times \bb \right) \times \delta \bb + \left(\nabla \times \delta \bb \right) \times \bb \right],
\end{equation}
can be found, and the frequency shifts \eqref{eq:freqpert} can be computed.


For the $f$-modes, we find a fit
\begin{equation} \label{eq:deltaomegaf}
\begin{aligned}
\frac{ \delta \omega^{B}_{f}} {50.3 \text{ Hz}} =& - \Bn^2 \Mn^{-3/2} \Rn^{5/2} \left( 1 - \frac {0.951} {\Lambda} \right) \\
&\times \Big( 1 - 2.15 \nun + 5.45 \nun^2 \\
& - 5.14 \nun^3 + 1.37 \nun^4 \Big),
\end{aligned}
\end{equation}
while for the leading-order $r$-modes, we have
\begin{equation} \label{eq:deltaomegar}
\begin{aligned}
\frac { \delta \omega^{B}_{r}} {55.1 \text{ Hz}} =& \Bn^2 \Mn^{-1} \Rn \left( 1- \frac {1.15} {\Lambda} \right) \\
&\times \left( 1 - 0.98 \nun^{-1} - 2.38 \nun + \nun^2 \right).
\end{aligned}
\end{equation}
Note that $\delta \omega^{B}_{r}$ scales with $\Omega^{-1}$ because $\omega_{r} \propto \Omega$ \eqref{eq:rmodefreq1} and an inverse power of $\omega_{r}$ is picked up from \eqref{eq:freqpert}.

\begin{figure}
\includegraphics[width=0.493\textwidth]{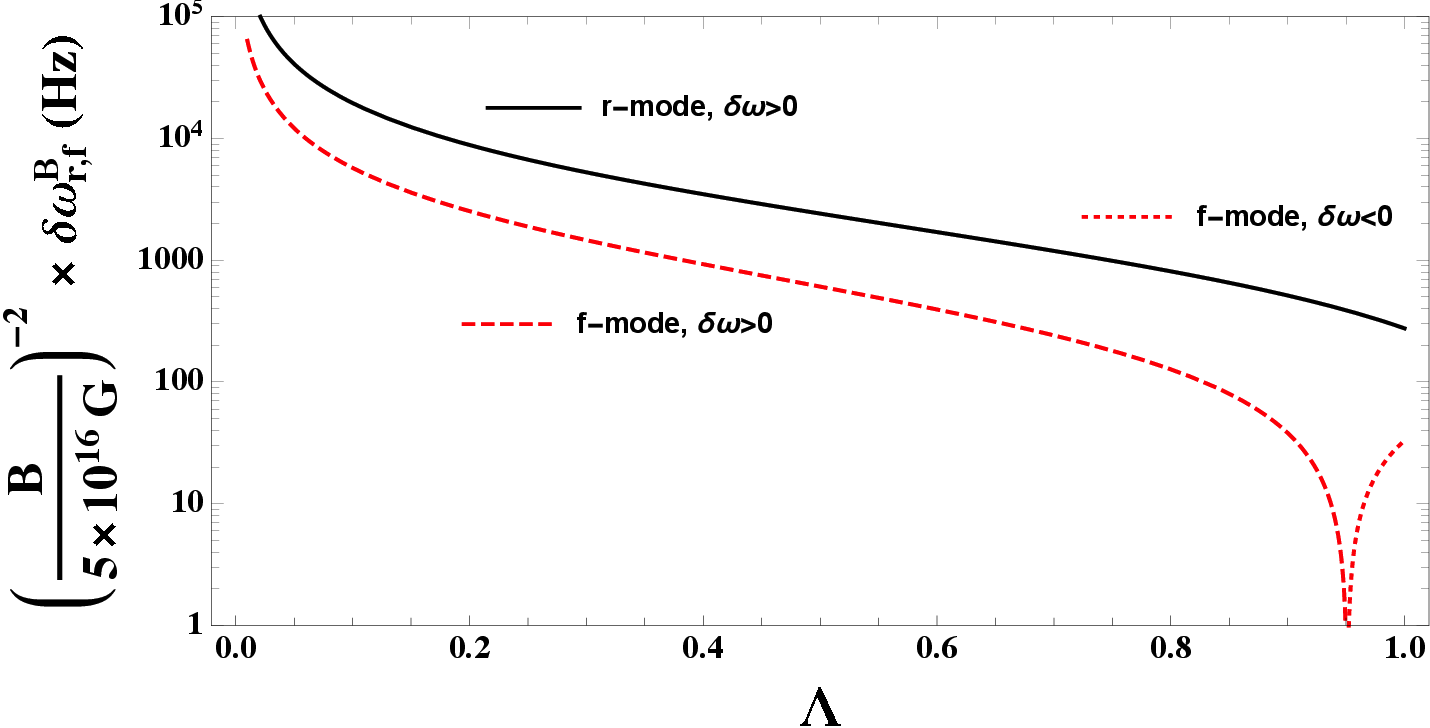}
\caption{Eigenfrequency shifts $\delta \omega^{B}$ for $f$- [red, dashed ($\delta \omega^{B}_{f} > 0$) and dotted ($\delta \omega^{B}_{f} < 0$) curves] and $r$- (black curve) modes due to the magnetic field as a function of the poloidal-to-toroidal field strength ratio $\Lambda$. We have taken $\nu = 300$ Hz, $M = 1.4 M_{\odot}$, and $R = 13 \text{ km}$. \label{deltaB}}
\end{figure}

In Figure \ref{deltaB} we plot the relative mode frequency shifts for the $f$- [red, dashed ($\delta \omega^{B}_{f} > 0$) and dotted ($\delta \omega^{B}_{f} < 0$) curves; expression \eqref{eq:deltaomegaf}] and $r$- [black curves; expression \eqref{eq:deltaomegar}] modes as functions of the poloidal-to-toroidal strength $\Lambda$ for some fixed stellar parameters. We see that the $r$-mode frequencies are more sensitive to the magnetic field in general, with $\delta \omega^{B}_{r} / \delta \omega^{B}_{f} \sim 4$ for $\Lambda \sim 0.5$. However, in any case, large field strengths $B \gtrsim 10^{16} \text{ G}$ are required to introduce a significant shift, unless the toroidal field is dominant \cite{lander10,lanjon11,asai15}. In this latter instance, the frequency shifts are always positive due to the sign of the Lorentz force \eqref{eq:lorentz}, which, in general, makes resonance more difficult to achieve for $f$-modes, though easier for $r$-modes since they have negative frequency in the inertial frame. The presence of a magnetic field also implies that Alfv{\'e}n modes can be excited \cite{sot09}, though we will not consider resonances with these modes here.

We caution the reader that, because we have neglected back-reaction effects into $\boldsymbol{\xi}$ from the perturbing forces, expression (35) gives us that $\delta \omega_{r} \propto \nu^{-1}$. As such, the expressions presented above for $r$-modes should only be considered valid for $\nu \gtrsim 30 \text{ Hz}$ {(cf. also the criterion $\nu \gtrsim 39 \text{ Hz}$ set by viscous damping \cite{bild99,lev01} discussed in Sec. III A)}, else the approximation scheme likely breaks down. {Since the magnetic force $\boldsymbol{F}^{B}$ is highly anisotropic in general (especially for non-dipole fields; see Footnote 6), the functional form of the eigenfunctions $\boldsymbol{\xi}$ may also be impacted if the field is ultra-strong \cite{lander10,lanjon11}. } 

Additionally, in presenting the computed shifts \eqref{eq:deltaomegaf} and \eqref{eq:deltaomegar}, we have taken rather large values for both the spin frequency and magnetic field simultaneously, which might be objectionable due to spin-down arguments. In particular, a magnetar with strong magnetic field is expected to spin slowly after several spin-down time-scales have elapsed, i.e. after several $t_{\text{sd}} \sim 2.2 \times 10^{2} \Bn^{-2} \nun^{-2}$ s post-birth. For a strong magnetic field and rapid rotation to co-exist, we require that: (i) the magnetar is young, or (ii) the magnetar has a surface magnetic field much weaker than the core field (cf. \cite{suvgep16,tiengo13}), or (iii) the magnetar resisted significant electromagnetic spin-down during the binary lifetime due to angular momentum accretion \cite{farr11} [cf. expression \eqref{eq:spinup}] .

\subsection{Total Shift}

Putting everything from the previous sections together, we thus have that
\begin{equation} \label{eq:total}
\omega = \omega_{0} + \delta \omega^{T}+  \delta \omega^{B}.
\end{equation}

\begin{figure*}
\includegraphics[width=\textwidth]{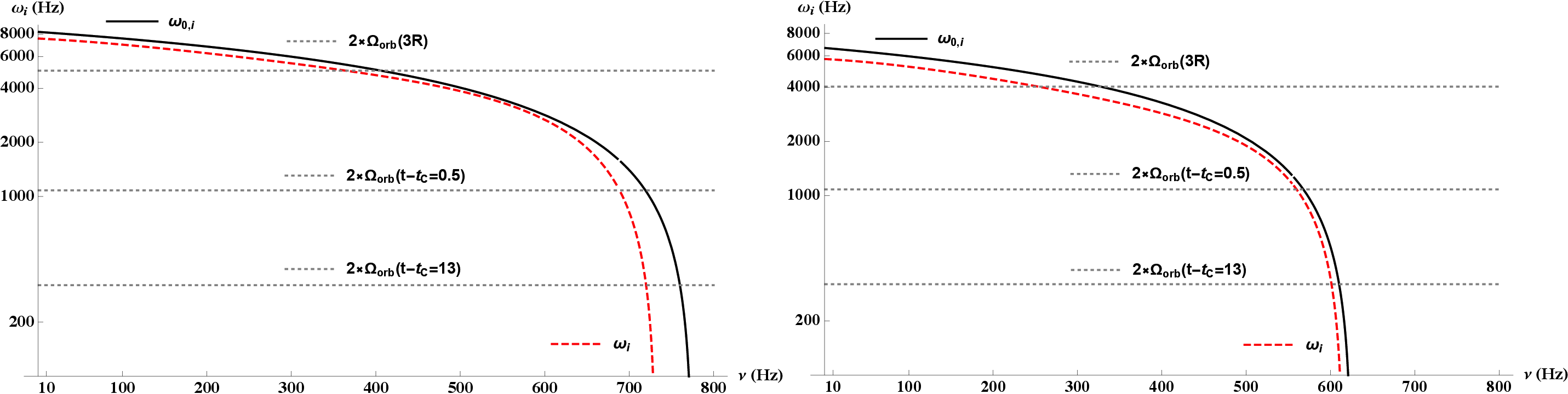}
\caption{Unperturbed ($\omega_{0,i}$; black curve) and perturbed (expression \eqref{eq:total}; red, dashed curve) $f$-mode frequencies in the inertial frame as a function of spin frequency $\nu$, where we set $R = 13 \text{ km}$ (left panel) and $R = 15 \text{ km}$ (right panel). Overplotted are twice the values of the orbital frequencies relevant for GRB 090510 (cf. Fig. \ref{orbsep}). \label{fMode}}
\end{figure*}

In Figure \ref{fMode} we show the unperturbed ($\omega_{0,i}$; black curve) and perturbed [expression \eqref{eq:total}; red, dashed curve] $f$-mode frequencies in the inertial frame as a function of spin frequency $\nu$ with $a=10^{7} \text{ cm}$, $M = \Mc = 1.4 M_{\odot}$, $B_{0} = 10^{17} \text{ G}$ for a purely poloidal configuration $(\Lambda =1)$, and mode amplitude $\alpha_{f} = 0.1$, with stellar radii $R = 13 \text{ km}$ (left panel) and $R = 15 \text{ km}$ (right panel). Over-plotted are the twice the values of the orbital frequency (the GW frequency times $2 \pi$) associated to the GRB 090510 precursors \cite{troja10}, together with the maximum GW frequency attainable prior to coalescence. Even for this extreme magnetic field strength, the mode frequency shifts are relatively modest, in agreement with results from the literature \cite{lander10,lanjon11,asai15}. 

For $R = 13 \text{ km}$, a rotation rate near the secular stability limit $\nu \approx 760 \text{ Hz}$ would be required for tidal resonances of the `unperturbed' $f$-modes with frequency $\omega_{0,i}$ to potentially explain precursor activity $\sim 13$ s prior to the merger, while $\nu \approx 740 \text{ Hz}$ is needed for the perturbed modes with frequency $\omega_{i}$ \cite{ho99}. Rotation rates required for precursors occurring closer to coalescence with $t-\tC \approx 0.5$ s are lower by $\approx 10\%$. For a larger stellar radius $R = 15 \text{ km}$ (possibly already excluded by GW170817 \cite{most18}), the necessary spin frequency drops to $\nu \sim 550 \text{ Hz}$, which is more reasonable (cf. $\nu = 716 \text{ Hz}$ for the fastest known pulsar PSR J$1748$-$2446$ad). We thus conclude that, unless the main burst occurs $\gtrsim 100$ ms later than the coalescence time $(t_{B} > \tC$; see Sec. II), which would imply that the associated orbital frequencies are larger than those presented in Tab. \ref{tab:sgrbdata} and thus are closer to the maximum frequency, it is difficult for $f$-mode resonance to explain early-time ($\tC - t \lesssim 10$ s) precursor activity, even when strong magnetic fields and tidal frequency shifts are considered. 

\begin{figure*}
\includegraphics[width=\textwidth]{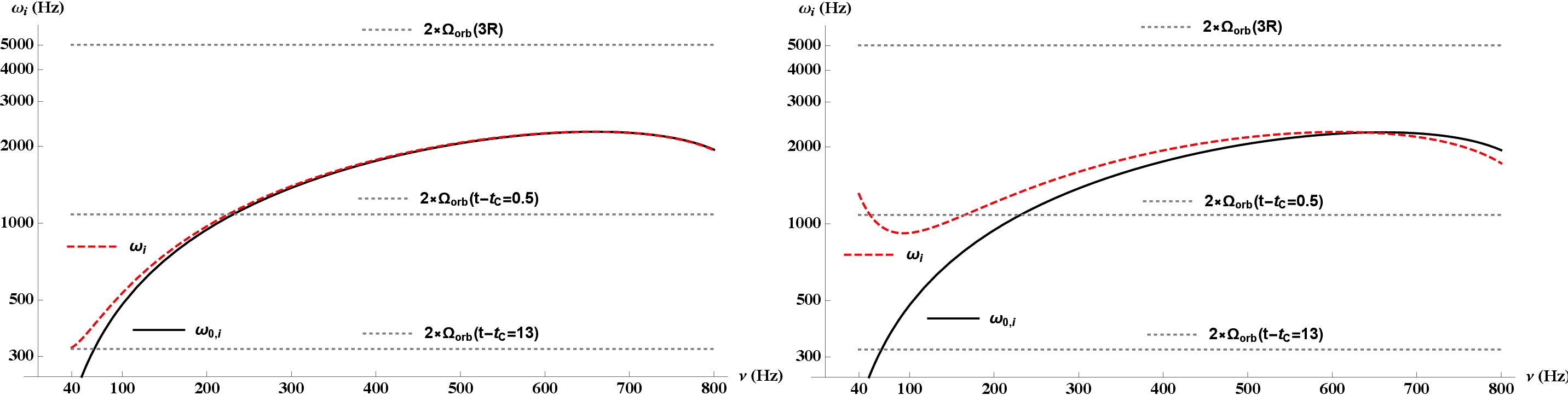}
\caption{Unperturbed ($\omega_{0,i}$; black curve) and perturbed (expression \eqref{eq:total}; red, dashed curve) $r$-mode frequencies in the inertial frame as a function of spin frequency $\nu$, where we set $\Lambda =1$ (left panel) and $\Lambda = 0.3$ (right panel). Overplotted are twice the values of the orbital frequencies relevant for GRB 090510 (cf. Fig. \ref{orbsep}). Note that $\omega_{i} < 0$ for these modes; they are subject to the gravitational radiation (CFS) instability \cite{fm78,and98,morsink98}. \label{rMode}}
\end{figure*}

Figure \ref{rMode} is similar to Fig. \ref{fMode} except it shows the $r$-mode frequencies, with the (somewhat) more modest values $B_{0} = 10^{16} \text{ G}$ and $R = 1.3 \times 10^{6} \text { cm}$, where we have a purely poloidal magnetic field $\Lambda =1$ (left panel) and a field with significant toroidal component $\Lambda = 0.3$ (right panel). In this case, low rotation rates are favoured to explain the precursors; for $\nu \sim 40 \text{ Hz}$ the resonance condition \eqref{eq:resonance} can be satisfied, with the exact intersection point shifting by a factor $\lesssim 2$ when frequency perturbations, dominated by the magnetic contribution \eqref{eq:deltaomegar}, are taken into account. Interestingly, since the toroidal field increases the mode frequency, even for $\Lambda = 0.3$ we see that early-time resonances are harder to achieve because the magnetic field leads to an inflection in the eigenfrequency profile at $\nu \sim 100 \text{ Hz}$. A weaker toroidal field is thus favoured to explain precursors in the tidal resonance scenario for both $f$- and $r$-modes. 

Assuming the resonant amplitudes are large enough (see below), two precursor events (i.e. in GRB 090510) might be explainable, for a NS with $\nu \lesssim 200 \text{ Hz}$, in the following way. An early-time resonance with the $r$-modes occurs some $\sim 10$ s prior to coalescence, triggering the first precursor. Then, assuming that our estimates for $\tC - t$ are high for the reasons discussed above, an $f$-mode resonance (or direct tidal shattering \cite{penner11}) might occur near the maximum GW frequency, giving rise to a second precursor. Additionally, although the growth time-scale is likely much longer than the resonance time-scale \eqref{eq:restime}, the CFS instability \cite{fm78,and98,morsink98} may increase $r$-mode amplitudes by up to factors $\gtrsim 10$ \cite{kokster98,rmodeamp1,rmodeamp2}. This would imply that spin-up due to resonance \eqref{eq:spinup} could be non-trivial, so that multiple $r$-mode resonances might occur if the $r$-mode amplitude increases significantly over $\sim 10$ s between precursors due to this instability. However, the viability of this latter possibility depends critically on the interplay between the $r$-mode growth time-scale and the (viscous) spin-up time-scale. 



\section{Energetics}
Having considered the relationship between mode frequencies and resonances, we turn now to the associated energetics of the precursor. In particular, the stellar pulsation modes introduce a Lagrangian displacement which shears the crust to a degree which depends on the mode amplitude $\alpha$, which reach maximum values \eqref{eq:modeamp} during a period of resonance \cite{lai94,gold94}. If the stresses exceed a critical threshold, determined by the crystalline properties of the crust \cite{chug10,chug18}, it is possible that the crust may yield and fracture \cite{lander15,suvkok19}. We explore the relationship between the amplitudes necessary to instigate crustal failure with the values obtained from tidal resonances \eqref{eq:modeamp}.


\subsection{Crustal failure}

In general, the elastic strain tensor $\boldsymbol{\sigma}$ has components (e.g. \cite{ll70})
\begin{equation} \label{eq:straindefn}
\sigma_{ij} = \frac {1} {2} \left( \nabla_{i} \xi_{j} + \nabla_{j} \xi_{i} \right).
\end{equation}

We assume that the crust fails when the von Mises criterion, coming from classical elasticity theory, is met. This happens when (see e.g. \cite{suvkok19})
\begin{equation} \label{eq:vonmises}
\sigma \equiv \sqrt{ \tfrac {1} {2} \sigma_{ij} \sigma^{ij} } \gtrsim \sm,
\end{equation}
where $\sm$ is the maximum breaking strain that the crust can sustain, and $\boldsymbol{\sigma}$ is the strain tensor defined in \eqref{eq:straindefn}. The recent semi-analytic lattice stability models of Chugunov \& Baiko \cite{chug18} find $\sm \approx 0.04$.

For the $f$-modes, using \eqref{eq:fmodexi} and \eqref{eq:straindefn}, we find the relationship
\begin{equation} \label{eq:fmodesigma}
\sigma_{f} = 2 \sqrt {2} \alpha_{f},
\end{equation}
which has no spatial dependence. This is due to the nature of the Maclaurin spheroid solutions, whereby the constant density and simple spatial profile for the Lagrangian displacement \eqref{eq:fmodexi} states that the crust is strained uniformly, so that it either fails either everywhere or nowhere. In this case, the volume of the failure region, if $\sigma_{f} \gtrsim \sm$, is just the crustal volume $V_{c} \approx \tfrac {4 \pi} {3} (R - R_{c})^3$, where $R_{c} \sim 0.9 R$. A more realistic equilibrium model (see e.g. \cite{pons11,lander15,suvkok19} and below) would give rise to some spatial dependence for $\sigma_{f}$. Nevertheless, we have from \eqref{eq:fmodesigma} that if
\begin{equation} \label{eq:fnecessary}
\alpha_{f} \gtrsim \frac {1} {50 \sqrt{2}} \left( \frac {\sm} {0.04} \right),
\end{equation}
then crustal failure would be expected. The resonant amplitude \eqref{eq:famp} exceeds the fracturing amplitude \eqref{eq:fnecessary} by an order of magnitude. 

For the $r$-modes, the crustal strain $\sigma$ can also be written down easily using the $r$-mode displacement \eqref{eq:rmodexi}, though the expression is lengthy so we avoid it here.

\begin{figure*}
\includegraphics[width=\textwidth]{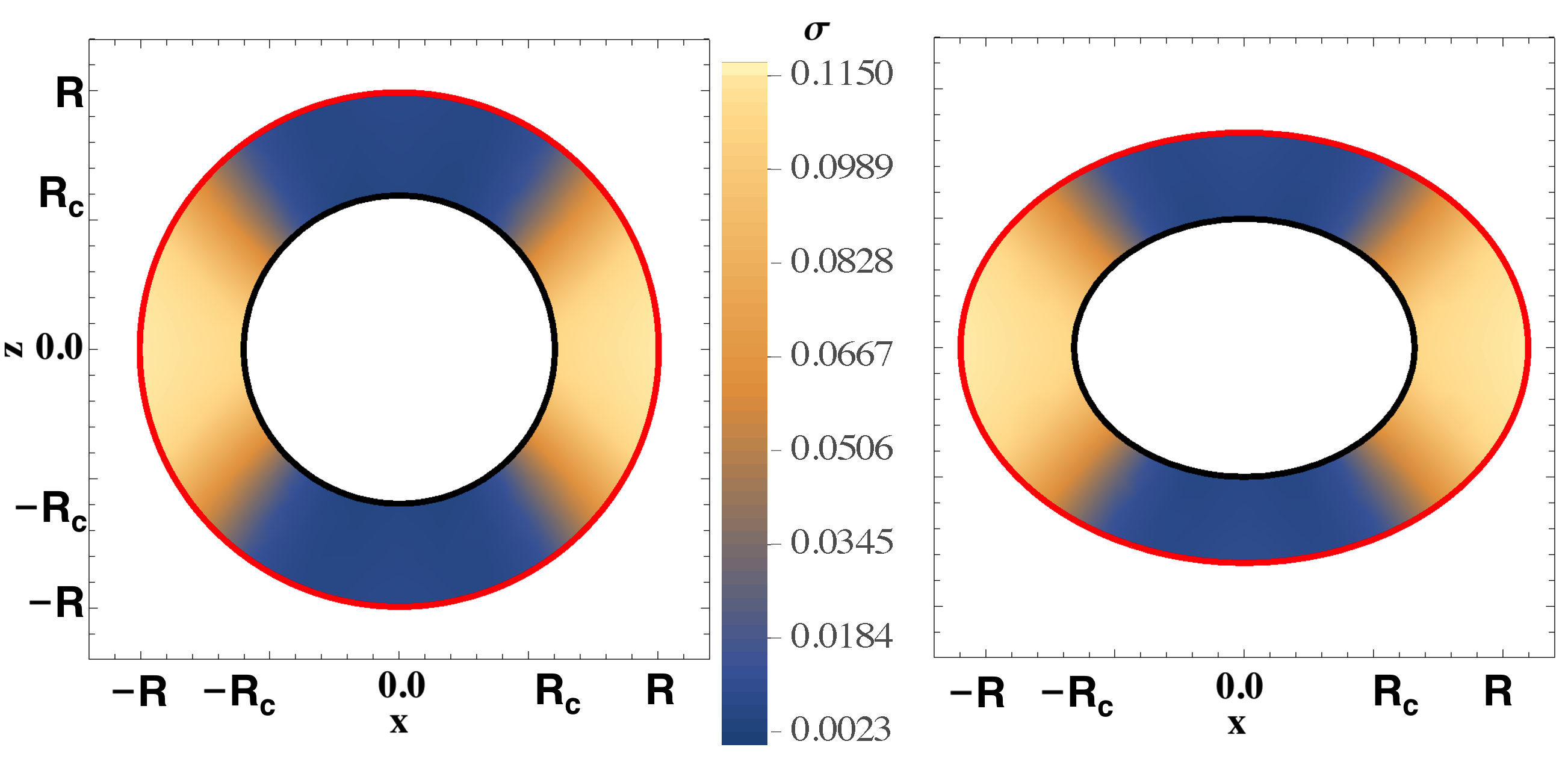}
\caption{Crustal strain $\sigma$ for the Lagrangian displacement induced by $r$-modes \eqref{eq:rmodexi} for $\nu = 100 \text{ Hz}$ (left panel) and $\nu = 500 \text{ Hz}$ (right panel), with amplitude $\alpha_{r} = 2 \times 10^{-4}$. Brighter shades indicate a greater value for $\sigma$. The black curve indicates the crust-core transition, while the red curve shows the (spin frequency dependent) stellar surface. (The crust has been stretched by a factor of $4$ for improved visibility). \label{rsigma}}
\end{figure*}

Figure \ref{rsigma} shows the $r$-mode crustal strain $\sigma$ for spin frequencies $\nu = 100 \text{ Hz}$ (left panel) and $\nu = 500 \text{ Hz}$ (right panel), where we set the mode amplitude $\alpha_{r} = 2 \times 10^{-4}$; a value which is an order of magnitude smaller than the resonant maximum \eqref{eq:ramp} for $\nu \sim 100 \text{ Hz}$. We see that rotation decreases the strain near the poles, while simultaneously increasing the strain near the equator, which is stronger by a factor $\sim 10^2$ in either case. The von Mises criterion \eqref{eq:vonmises} with $\sm = 0.04$ suggests that the crust would yield only in equatorial regions for the resonant amplitude \eqref{eq:ramp} if $\nu \lesssim 100 \text{ Hz}$, though would yield almost everywhere if $\nu \gtrsim 500 \text{ Hz}$. Note also that we have taken $\omega = \omega_{0}$ when computing $\sigma$, though the corrections discussed in Sec. IV will impact the geometry of the failure region if, for example, the magnetic field is strong $(B \gtrsim 10^{16} \text{ G})$, especially for low spin frequencies.


\subsection{Crustal fracture energy}

In the previous section, we found that the von Mises criterion \eqref{eq:vonmises} can be met for both $f$- and $r$-modes. Here we adopt the simple prescription that crustal failure implies a quake and energy release, though the situation is likely to be more complicated in reality. Depending on thermodynamic aspects of the crust, such as the melting temperature, phases of plastic flow may be induced rather than fracturing \cite{jones03,belo14}.

We wish to estimate the amount of energy potentially released due to crustal fracturing, to see if it is consistent with that which is observed from the precursor flares. To do this, we consider the available energy density $U$ integrated over the resonance interval $[0, t_{\text{res}}]$, where $t_{\text{res}}$ is given by \eqref{eq:restime}, and over regions where the crust actually yields, i.e. in regions where the von Mises criterion \eqref{eq:vonmises} is met. As such, we write
\begin{equation} \label{eq:energyeqn}
\begin{aligned}
\int dt E_{\text{quake}} &= \int^{t_{\text{res}}}_{0} dt \int_{\sigma \geq \sm} dV  U(t,\boldsymbol{x}) \\
&\approx t_{\text{res}} \int_{\sigma \geq \sm} dV  U(0,\boldsymbol{x}),
\end{aligned}
\end{equation}
is the energy released, during a period of tidal resonance, as a result of crustal activity.  Expression \eqref{eq:energyeqn} qualitatively agrees with that of Ref. \cite{lander15} (though these authors considered magnetic stresses). The major contributions to $U$ come from the rotational $U_{\text{rot}} = \tfrac{1}{2} \rho \bv \cdot \bv$ and magnetic  $U_{\text{mag}} = \left( 8 \pi \right)^{-1} \boldsymbol{B} \cdot \boldsymbol{B}$ energy densities. There is also a tidal contribution, though it is negligible except in the final  stages of inspiral $a \lesssim 5 R$. In particular, expression \eqref{eq:energyeqn} then gives us that
\begin{equation} \label{eq:quakeenergy}
\begin{aligned}
\int dt E_{\text{quake}} \approx& \,\, 1.64 \times 10^{47} \text{ erg s} \\
& \times \left( \frac {t_{\text{res}}} {3.6 \times 10^{-2} \text{ s}} \right) \times \left[ \frac {\text{Vol}(\sigma \geq \sm)} {5 \times 10^{15} \text{ cm}^3} \right] \\
& \times \left( \Mn \nun^2 \Rn^{-1}  + 4.4 \times 10^{-3} \Bn^2 \right) ,
\end{aligned}
\end{equation}
which should be compared with \eqref{eq:maxL}. We may conclude therefore that crust yielding due to $f$- and $r$-mode resonances, if the amplitudes $\alpha$ reach the resonant values \eqref{eq:modeamp}, can accommodate, energetically speaking, SGRB precursor events.

\section{Discussion}

In NSNS binaries, mutual tidal interactions prior to coalescence can naturally lead to the excitation of stellar oscillation modes. If the oscillations come into resonance with the orbital motion \eqref{eq:resonance}, tidal energy is rapidly absorbed by the mode(s) over some resonance time-scale \eqref{eq:restime} \cite{alex87,lai94,kokk95}, and the respective amplitudes grow substantially (Sec. III. B). If the resonant amplitudes are large enough (see Fig. \ref{rsigma}), the shear stresses $\sigma$ exerted on the crust as a result of the (magneto-)hydrodynamic displacement $\boldsymbol{\xi}$ can exceed some critical threshold \eqref{eq:vonmises}, causing a crustal failure event, such as a quake. In particular, there is strong evidence that SGRB events are associated with binary NS mergers \cite{sgrb1,sgrb2,sgrb3}, and some bursts (see Tab. \ref{tab:sgrbdata}) are known to be preceded by episodes of early emission (`precursor flares'). In light of the above, it has been suggested that crustal failures, instigated by tidal resonances prior to coalescence, may source the SGRB precursors \cite{tsang12,tsang13}. In this paper we further develop this model by including magnetic fields (noting that there is some evidence to suggest that binaries exhibiting precursors contain magnetars; see Sec. II. C), and by comparing the resonant amplitude(s) with the crustal breaking strain for both $f$- and $r$-modes, to see what stellar parameters would be necessary to accommodate precursor data \cite{troja10,min17,zhong19}. As argued in Sec. IV. C, we found that multiple precursors from the same object (such as those seen in GRB 090510) might be accommodated by an early-time ($t_{B} - t \sim 10$ s) $r$-mode resonance, followed by a late-time $f$-mode resonance (or direct tidal shattering \cite{penner11}) $\gtrsim 10^{2}$ ms prior to the merger.

In this work, we have only considered $f$- and $r$- mode resonances. However, $g$-modes (including torsional shear modes and interface modes \cite{hansen88}), which are perturbations restored by buoyancy {or are otherwise supported by a fluid-to-solid transition}, {may also be} important\footnote{{However, the presence of a solid crust tends to `squeeze' $g$-modes into the core \cite{hansen88}, which may impact their ability to exert strains.}}  in the precursor scenario, since they have typical frequencies $\lesssim 10^{2} \text{ Hz}$ in the co-rotating frame \cite{krug14} (see also Ref. \cite{pgmode}) and can have non-trivial overlap integrals \cite{lai94,kokk95,tsang12}. The $g$-mode frequencies are comparable with the $r$-mode frequencies for low rotation rates, and thus offer an alternate avenue for explaining multiple precursors from tidal resonances, i.e. one could imagine two flares being sourced via crust yielding through separate instances of $g$- and $r$- mode resonances. Furthermore, in addition to transferring energy \eqref{eq:energytransfer} and angular momentum \eqref{eq:spinup}, tidal interactions lead to heating through friction, and can raise the NS temperature to $\gtrsim 10^{8}$ K prior to coalescence \cite{lai94}. As such, buoyancy-driven modes arising due to thermal gradients may be especially significant in the resonance scenario.  Additionally, if hot enough, the NSs may shed mass through a radiation-driven wind prior to merging at $a \lesssim 3 R$ \cite{rees92}, which can lead to baryon loading in the region which comes to surround the post-merger remnant. This would reduce the $\gamma$-ray luminosity of the main GRB episode due to absorption. Unfortunately, $g$-modes cannot be accommodated by Maclaurin spheroid models because there are no compositional gradients to support a non-zero Brunt-V{\"a}is{\"a}l{\"a} frequency. A thorough exploration of tidally-forced $g$-mode properties and tidal heating for magnetised neutron stars using more realistic stellar models will be conducted elsewhere.

In addition to the absence of compositional gradients, it is important to note that several approximations are present in the simple, analytic calculations presented here. Though it has been found that $f$- and $r$-mode frequencies for Maclaurin spheroids reasonably approximate the values obtained when using realistic EOS (see Fig. \ref{fAndrFreq}), the elastic response of the crust \eqref{eq:straindefn} is sensitive to the exact profile of the displacement vector, and so our models should not be taken as representative of real astrophysical NSs. Detailed analyses using sophisticated simulations of the NS quasi-normal mode spectrum, involving realistic equations of state \cite{kru19} and tidal couplings/deformabilities \cite{hind10,sterg19}, would be useful in this direction.

\section*{Acknowledgements}
This work was supported by the Alexander von Humboldt Foundation, DFG research grant 413873357, and the DAAD program ``Hochschulpartnerschaften mit Griechenland 2016'' (Projekt 57340132). {We thank the anonymous referee for their valuable feedback, which improved the quality of the manuscript.} AGS thanks Nick Stergioulas for hospitality shown at the University of Thessaloniki where some of this work was completed.



\end{document}